\documentclass[
    ,final            
  ]
  {aipproc}

\layoutstyle{6x9}

\usepackage{latexsym}
\bibpunct{(}{)}{;}{a}{}{,}

\def\hide#1{}

\def \apj {ApJ}

\def \apjl {ApJL}
\def \aap {A\&A}
\def \nat {Nature}

\def \iaucirc {IAUC}

\begin{document}

\title{Timing the accretion flow around accreting millisecond pulsars}

\classification{97.60.Jd 97.80.Jp 97.60.Gb}
\keywords      {Neutron stars - X-ray binaries - Pulsars}

\author{Manuel Linares}{
  address={Astronomical Institute ``Anton Pannekoek'', University of Amsterdam, Kruislaan 403, NL-1098 SJ Amsterdam, Netherlands. mail to: linares@uva.nl}
}



\begin{abstract}

At present, ten years after they were first discovered, ten accreting
millisecond pulsars are known. I present a study of the aperiodic
X-ray variability in three of these systems, which led to the
discovery of simultaneous kHz quasi periodic oscillations in
XTE~J1807--294 and extremely strong broadband noise at unusually low
variability frequencies in IGR~J00291+5934. Furthermore, we classified
SWIFT~J1756.9--2508 as an atoll source and measured in its 2007
outburst spectral and variability properties typical of the extreme
island state. I also give detailed estimates of the total fluence
during the studied outbursts.

\end{abstract}

\maketitle


\section{Introduction}

During the last decade accreting millisecond pulsars (AMPs) have
revealed a number of interesting phenomena and have opened a new
window to the physics of accretion onto neutron stars (NSs). The first
of such systems was discovered by
\citet[][SAX~J1808.4--3658]{Wijnands98}, presenting the first prove of
an accreting neutron star having both millisecond spin period and
dynamically important magnetic field. Ten AMPs have been discovered to
date, and in three of them millisecond X-ray pulsations have been seen
to appear in and disappear from the persistent emission, producing
predominant \citep[HETE~J1900.1-2455;][]{Galloway07} intermittent
\citep[SAX~J1748.9-2021;][]{Altamirano07} or very rare
\citep[Aql~X-1;][]{Casella07} episodes of pulsations. This implies that
the AMP within them is only active or visible during a relatively
small fraction of the time, which may provide a link with the much
more numerous class of non-pulsating neutron star low-mass
X-ray binaries (NS-LMXBs).

One way to study accretion onto compact objects is to analyze the
aperiodic variability in the X-ray flux coming from these sources,
which tells us about processes occuring in the inner accretion flow
\citep{vanderKlis95,vanderKlis06}. Such timing of the accretion flow,
combined with a study of the X-ray spectrum, reveals different
``accretion states''. We show in this paper three different AMPs in
three different accretion states, we describe their aperiodic variability
and quantify their outburst fluence.

\section{IGR~J00291+5934}

The sixth AMP was discovered on December $2^{nd}$, 2004:
IGR~J00291+5934 \citep{Eckert04}. Coherent pulsations were found at a
frequency of 598.9~Hz, modulated by the $\sim$2.5~hr orbital motion
\citep{Markwardt04b,Markwardt04c}. Follow up observations of the
outburst, which lasted about two weeks, were performed by {\it RXTE}
(see Fig.~2). Our study of IGR~J00291+5934 showed what still constitutes
the strongest X-ray variability seen in a NS-LMXB, namely the
fractional rms was $\sim$50\%. We also measured the lowest
charactersitic frequencies ever seen in a NS-LMXB, with a break in the
flat-topped power spectrum at $\sim$0.04~Hz \citep[see Figure~1
and ref.][for more details and discussion]{Linares07}. This AMP is
therefore an extreme example of what is often called ``extreme
island'' state of atoll sources \citep[a low-luminosity class of
NS-LMXB, see][]{HK89}.

\begin{figure}[h]
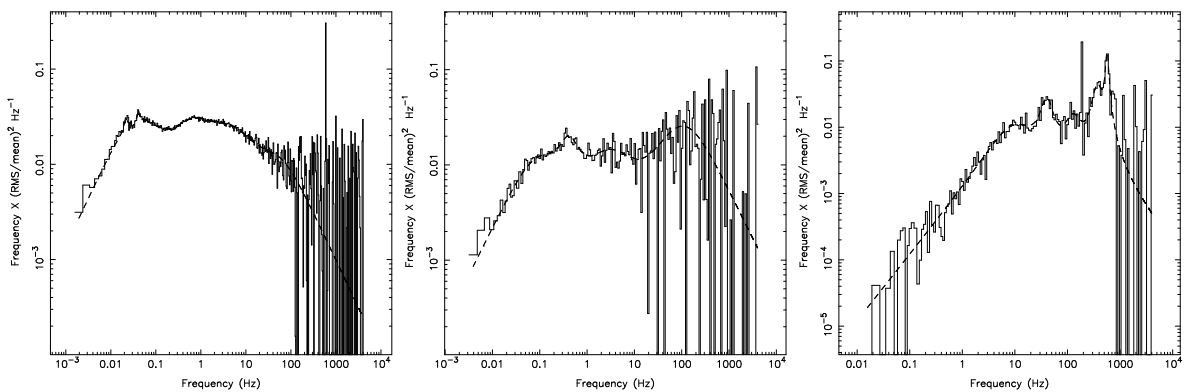

\includegraphics[height=.23\textheight]{igrj00291_A+B.ps}
\includegraphics[height=.23\textheight]{swj1756_A+B.ps}
\includegraphics[height=.23\textheight]{xtej1807_H.ps} 
\caption{Power spectra (in power times frequency representation and rms normalized)
of three AMPs in outburst: IGR~J00291+5934 {\it (left)}, SWIFT~J1756.9--2508
{\it (center)} and XTE~J1807--294 {\it (right)}. The break frequency in IGR~J00291+5934 is
more than two orders of magnitude lower than that of XTE~J1807--294,
which shows instead simultaneous kHz QPOs.}
\end{figure}

\section{SWIFT~J1756.9--2508}

On June $7^{th}$, 2007, a new X-ray transient was discovered
\citep{Krimm07a} with the burst alert telescope (BAT)
onboard {\it Swift}. Follow up {\it RXTE} observations revealed that
this was the eighth discovered AMP and showed a pulse frequency of
$\sim$182~Hz and an orbital period of $\sim$54~minutes
\citep{Markwardt07a,Markwardt07b, Krimm07c}. The outburst lasted about two
weeks (see Fig.~2). We analyzed the aperiodic variability of the
source, comparing it with other AMPs and with atoll sources. We
thereby classified SWIFT~J1756.9-2508 as an atoll source in the
extreme island state. Using both PCA and HEXTE data we detected a hard
tail in its energy spectrum extending up to 100~keV, fully consistent
with such source and state classification \citep{Linares08a}. It is
interesting to note that so far all AMPs show spectral and timing
\citep[except for the shifts in the frequency-frequency correlations,
see][]{Straaten05,Linares05} properties identical to those of atoll
sources, which suggests that low mass accretion rate is a necessary
(even though seemingly not sufficient) ingredient to make an AMP.

\section{XTE~J1807--294}

A new transient X-ray source was discovered in the Galactic bulge
region on February $13^{th}$, 2003. Subsequently, coherent pulsations
were detected at a frequency of 190.6~Hz turning the new system,
XTE~J1807--294, into the fourth discovered AMP
\citep{Markwardt03c}. An orbital period of $\sim$40 minutes was
determined \citep{Markwardt03a}, still the shortest among AMPs. The
outburst was followed by {\it RXTE} during five months (see Fig.2). We
discovered seven pairs of twin kHz quasi-periodic oscillations in
XTE~J1807--294 \citep{Linares05}, with a frequency separation
approximately equal to the spin frequency (see M. van der Klis
contribution in these proceedings for further details).

\begin{figure}[h]
  \includegraphics[height=.17\textheight]{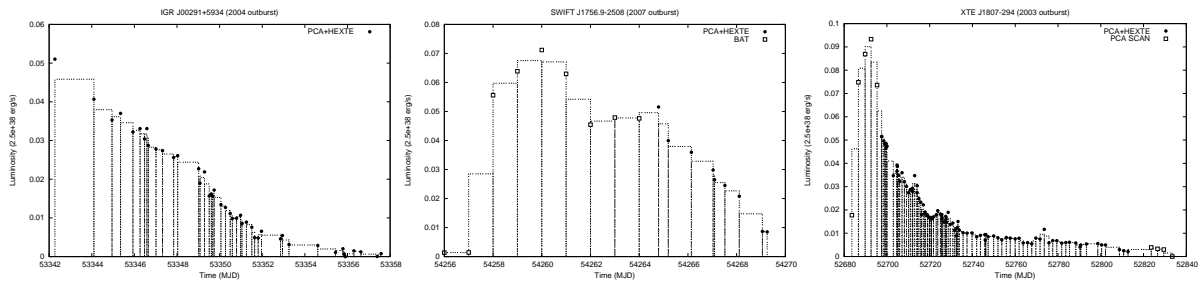}
  \caption{Light curves of the three AMP outbursts studied herein. Filled circles show the {\it RXTE}-PCA\&HEXTE measurements and empty squares those of {\it Swift}-BAT or PCA scans.}
\end{figure}

\section{Outburst fluences}

An important question that still remains open is why AMPs show
pulsations whereas most of NS-LMXBs do not. The solution proposed by
\citet{Cumming01} invokes screening of the magnetic field in
``classical'' NS-LMXBs by a time-averaged mass accretion rate higher
than that of AMPs. In order to test this and other theories (NS crust
cooling, binary evolution) a careful estimate of how much mass falls
onto the neutron star surface is of capital importance
\citep{Galloway06b}. For this purpose we measure the unabsorbed
2-200~keV flux during the three AMP outbursts mentioned above, using
data from both PCA and HEXTE onboard {\it RXTE}. We fit the broadband,
background and deadtime corrected, energy spectra with an absorbed
disk blackbody plus power law model (fixing the column density to the
Galactic value in the source direction). For XTE~J1807--294 we use
data from the PCA scans of the Galactic bulge \citep{Markwardt02} to
cover the rise and the final decay, as there are no pointed {\it RXTE}
observations during those parts of the outburst. In the case of
SWIFT~J1756.9--2508 we use data from the {\it Swift}-BAT transient
monitor\footnote{\url{http://swift.gsfc.nasa.gov/docs/swift/results/transients/}},
as most of the outburst had no {\it RXTE} pointings. We calibrate the
conversion between PCA-scan/BAT and PCA\&HEXTE fluxes in those parts
of the outbursts where both fluxes are available, thus implicitly
assuming that the spectral shape does not vary drastically. The
resulting lightcurves, translated to luminosities in units of
2.5$\times 10^{38}$ erg/s, are shown in Figure~2. Table~1 shows our
measurements of the fluence, radiated energy and peak luminosity for
the three outbursts studied (see note on distances used therein). We
also show in Table~1 the range spanned by the break frequency in the
power spectrum, which indicates the relative change in X-ray
variability frequencies, as well as the orbital period of each system.

\begin{table}[h!!]
\begin{tabular}{lccccc}
\hline
 & Outburst  &  Outburst  &  Peak & Break & Orbital \\
 & fluence & energy\tablenote{The fiducial distance used is indicated, in kiloparsecs and between brackets. These three AMPs have not shown X-ray bursts and their distances are therefore uncertain \citep[see, however, ][]{Falanga05,Galloway05}.} & luminosity$^*$ & frequency & period \\
 & ($10^{-3}$ erg/cm$^2$) &  ($10^{43}$ erg) & (\% Eddington) & (Hz) & (min) \\
\hline
XTE~J1807--294 & 7.2 & 5.5 [8] & 9.5 [8] & 5.3-10.2 & 40 \\
IGR~J00291+5934 & 1.4 & 0.6 [6] & 5 [6] & 0.03-0.05 & 150 \\
SWIFT~J1756.9--2508 & 1.5 & 1.2 [8] & 7 [8] & 0.09-0.12 & 54 \\
\hline
\end{tabular}
\caption{Parameters of the AMP outbursts studied in this work.}
\label{tab:a}
\end{table}

\begin{theacknowledgments}

I thank my colleagues and members of the organizing committee, as well
as all the participants of the Amsterdam workshop, for making this
event a fruitful and enjoyable one.

\end{theacknowledgments}



\bibliographystyle{aipprocl} 



\begin{thebibliography}{26}
\expandafter\ifx\csname natexlab\endcsname\relax\def\natexlab#1{#1}\fi
\providecommand{\enquote}[1]{``#1''}
\expandafter\ifx\csname url\endcsname\relax
  \def\url#1{\texttt{#1}}\fi
\expandafter\ifx\csname urlprefix\endcsname\relax\def\urlprefix{URL }\fi
\providecommand{\eprint}[2][]{\url{#2}}


\bibitem[{Wijnands} and {van der Klis}(1998)]{Wijnands98}
R.~{Wijnands}, and M.~{van der Klis}, \emph{\nat} \textbf{394}, 344--346
  (1998).

\bibitem[{Galloway} et~al.(2007)]{Galloway07}
D.~K. {Galloway} et al., \emph{\apjl} \textbf{654}, L73--L76 (2007).

\bibitem[{Altamirano} et~al.(2008)]{Altamirano07}
D.~{Altamirano}, P.~{Casella}, A.~{Patruno}, R.~{Wijnands}, and M.~{van der
  Klis}, \emph{\apjl} \textbf{674}, L45--L48 (2008).

\bibitem[{Casella} et~al.(2008)]{Casella07}
P.~{Casella}, D.~{Altamirano}, A.~{Patruno}, R.~{Wijnands}, and M.~{van der
  Klis}, \emph{\apjl} \textbf{674}, L41--L44 (2008).

\bibitem[{van der Klis}(1995)]{vanderKlis95}
M.~{van der Klis}, \emph{{in "X-ray binaries"}},Cambridge University Press, eds. Lewin, W.H.G. et al., 1995.

\bibitem[{van der Klis}(2006)]{vanderKlis06}
M.~{van der Klis}, \emph{in "Compact Stellar X-ray Sources", ed. W. H. G. Lewin
  \& M. van der Klis (Cambridge Univ. Press)} pp. 39--112
  (2006).

\bibitem[{Eckert} et~al.(2004)]{Eckert04}
D.~{Eckert} et al., \emph{The Astronomer's
  Telegram} \textbf{352}, 1--+ (2004).

\bibitem[{Markwardt} et~al.(2004{\natexlab{a}})]{Markwardt04b}
C.~B. {Markwardt}, J.~H. {Swank}, and T.~E. {Strohmayer}, \emph{The
  Astronomer's Telegram} \textbf{353}, 1--+ (2004{\natexlab{a}}).

\bibitem[{Markwardt} et~al.(2004{\natexlab{b}})]{Markwardt04c}
C.~B. {Markwardt} et al., \emph{The Astronomer's Telegram} \textbf{360}, 1--+
  (2004{\natexlab{b}}).

\bibitem[{Linares} et~al.(2007)]{Linares07}
M.~{Linares}, M.~{van der Klis}, and R.~{Wijnands}, \emph{\apj} \textbf{660},
  595--604 (2007).

\bibitem[{Hasinger} and {van der Klis}(1989)]{HK89}
G.~{Hasinger}, and M.~{van der Klis}, \emph{\aap} \textbf{225}, 79--96 (1989).

\bibitem[{Krimm} et~al.(2007{\natexlab{a}})]{Krimm07a}
H.~A. {Krimm} et al., \emph{The Astronomer's
  Telegram} \textbf{1105}, 1--+ (2007{\natexlab{a}}).

\bibitem[{Markwardt} et~al.(2007)]{Markwardt07a}
C.~B. {Markwardt}, H.~A. {Krimm}, and J.~H. {Swank}, \emph{The Astronomer's
  Telegram} \textbf{1108}, 1--+ (2007).

\bibitem[{Markwardt} and {Krimm}(2007)]{Markwardt07b}
C.~B. {Markwardt}, and H.~A. {Krimm}, \emph{The Astronomer's Telegram}
  \textbf{1114}, 1--+ (2007).

\bibitem[{Krimm} et~al.(2007{\natexlab{b}})]{Krimm07c}
H.~A. {Krimm} et al.,
  \emph{\apjl} \textbf{668}, L147--L150 (2007{\natexlab{b}}),
.

\bibitem[{Linares} et~al.(2008)]{Linares08a}
M.~{Linares} et al., \emph{\apj} \textbf{677}, 515--519 (2008).

\bibitem[{van Straaten} et~al.(2005)]{Straaten05}
S.~{van Straaten}, M.~{van der Klis}, and R.~{Wijnands}, \emph{\apj}
  \textbf{619}, 455--482 (2005).

\bibitem[{Linares} et~al.(2005)]{Linares05}
M.~{Linares}, M.~{van der Klis}, D.~{Altamirano}, and C.~B. {Markwardt},
  \emph{\apj} \textbf{634}, 1250--1260 (2005).

\bibitem[{Markwardt} et~al.(2003{\natexlab{a}})]{Markwardt03c}
C.~B. {Markwardt}, E.~{Smith}, and J.~H. {Swank}, \emph{\iaucirc}
  \textbf{8080}, 2--+ (2003{\natexlab{a}}).

\bibitem[{Markwardt} et~al.(2003{\natexlab{b}})]{Markwardt03a}
C.~B. {Markwardt}, M.~{Juda}, and J.~H. {Swank}, \emph{\iaucirc} \textbf{8095},
  2--+ (2003{\natexlab{b}}).

\bibitem[{Cumming} et~al.(2001)]{Cumming01}
A.~{Cumming}, E.~{Zweibel}, and L.~{Bildsten}, \emph{\apj} \textbf{557},
  958--966 (2001).

\bibitem[{Galloway}(2006)]{Galloway06b}
D.~K. {Galloway}, edited by
  F.~{D'Amico} et al., 2006, vol. 840 of
  \emph{AIP Conference Series}, pp. 50--54.

\bibitem[{Markwardt} and {Swank}(2002)]{Markwardt02}
C.~B. {Markwardt}, and F.~J. {Swank}, \emph{APS APRX}, 11007 (2002).

\bibitem[{Falanga} et~al.(2005)]{Falanga05}
M.~{Falanga}, J.~M. {Bonnet-Bidaud}, J.~{Poutanen}, R.~{Farinelli},
  A.~{Martocchia}, and P.~{Goldoni} (2005).

\bibitem[{Galloway} et~al.(2005)]{Galloway05}
D.~K. {Galloway} et al., \emph{\apjl} \textbf{622}, L45--L48 (2005).


\end{thebibliography}


\end{document}